\documentclass[reprint,floatfix,prl,amsmath,amssymb]{revtex4-2}
\usepackage[utf8]{inputenc}
\usepackage{graphicx}
\usepackage{xcolor}
\usepackage[colorlinks,citecolor=violet,linkcolor=violet,urlcolor=violet]{hyperref}

\begin{document}

\title{Message passing and cyclicity transition}
\author{Takayuki Hiraoka}
\email{takayuki.hiraoka@aalto.fi}
\affiliation{Department of Computer Science, Aalto University, 00076 Espoo, Finland}

\begin{abstract}
Message passing, also known as belief propagation, is a versatile framework for analyzing models defined on graphs. Its most prototypical application is percolation; yet, the interpretation of the message passing formulation of percolation remains elusive. We show that the message passing solutions commonly associated with the probability of belonging to the giant component actually identify reachability from cycles. This interpretation generally applies to bond and site percolation on any directed or undirected networks. Our findings highlight the distinction between transition in cyclicity and the emergence of the giant component.
\end{abstract}

\maketitle

Message passing, also called belief propagation, is a simple and scalable algorithmic framework for solving models defined on networks. Originally developed in the field of machine learning and information theory, it has found broad applications to problems in statistical physics and network science, including graph spectra~\cite{rogers2008cavity,newman2019spectra,cantwell2023heterogeneous}, epidemic models~\cite{karrer2010message,altarelli2014bayesian,altarelli2014containing}, community detection~\cite{decelle2011inference,krzakala2013spectral,zhang2014scalable}, spin models~\cite{mezard2009information,yoon2011beliefpropagation,mezard2017meanfield,kirkley2021belief,wang2024tensor}, and path optimization~\cite{yeung2012competition,yeung2013physics}, among others~\cite{zdeborova2006number,altarelli2011stochastic,liu2011controllability,cantwell2022belief}. 

In message passing, each node sends messages that encode information about its state. The algorithm computes these messages based on local dependencies, formalized by message passing equations that describe how incoming messages are aggregated at each node and passed on to another. 
Perhaps the simplest message passing equation is:
\begin{equation}
x_{j \to i} = \prod_{k \in \partial_{j \to i}} x_{k \to j}.
\label{eq:mp_simple}
\end{equation}
Here, $x_{j \to i}$ denotes the message sent from node $j$ to node $i$. 
Equation~\eqref{eq:mp_simple} specifies that the messages received by node $j$ are aggregated via a product and sent to node $i$. To prevent the message from being echoed back, the product runs over $\partial_{j \to i} = \partial_j \setminus \{i\}$, where $\partial_i$ denotes the set of node $i$'s neighbors (i.e., nodes that can directly influence the state of $i$). The message passing equations, defined for every pair of nodes and their neighbors, constitute a self-consistent relation for the set of messages.

The purpose of this paper is to clarify the meaning of the message $x_{j \to i}$ in Eq.~\eqref{eq:mp_simple}. The prevailing understanding is that it denotes the probability that node $j$ does not belong to the giant component---the only extensive component that occupies a non-vanishing fraction of the network in the large size limit---in the absence of node $i$~\cite{shiraki2010cavity,newman2023message}. Node $j$ is in the giant component in the absence of node $i$ 
if and only if at least one neighbor of $j$ is also in the giant component without the involvement of $i$ or $j$.
Based on this premise, the message passing algorithm has been widely adopted to study percolation under the random removal of nodes and edges~\cite{hamilton2014tight,karrer2014percolation,radicchi2015breaking,kuhn2017heterogeneous,bianconi2017fluctuations,cantwell2019message}. 
Namely, the message passing equation is formulated as
\begin{equation}
x_{j \to i} = \prod_{k \in \partial_{j \to i}} [1 - q + q x_{k \to j}]
\label{eq:mp_bond}
\end{equation}
for bond percolation where each edge is retained independently with probability $q$, and 
\begin{equation}
x_{j \to i} = 1 - q + q \prod_{k \in \partial_{j \to i}} x_{k \to j}
\label{eq:mp_site}
\end{equation}
for site percolation where each node is retained independently with probability $q$. When the algorithm is applied to directed networks, the message $x_{j \to i}$ is typically interpreted as the probability that $j$ is reachable from the giant strongly connected component \cite{timar2017mapping}. 

Although this interpretation appears plausible, it warrants careful inspection. These equations can be solved by setting a random initial value in $(0, 1)$ for each message and iterating the equation until convergence. Previous studies benchmarking the message passing method for estimating the percolation threshold and the size of the largest component have yielded mixed results: While the algorithm generally shows good agreement with Monte Carlo simulations for many synthetic and real-world networks, it performs poorly in some cases~\cite{faqeeh2015network,radicchi2016locally,timar2017nonbacktracking,allard2019accuracy,pastor-satorras2020localization,kim2026belief}. 

Beyond accuracy issues, the conceptual basis of this interpretation is not entirely clear. It is apparent from the construction of these equations that they describe the reachability of each node from something. Yet, nothing in the formulation explicitly proclaims that the upstream object in question is the giant component. Moreover, these equations can be formulated and solved for any finite network, including the ones for which the notion of a giant component may be ill-defined. In such a case, how does the algorithm know whether the component is extensive or not?

Studies have tried to address this puzzle by translating the original finite network into an infinite network via a process called \emph{unwrapping} or \emph{cloning}~\cite{hamilton2014tight,faqeeh2015network,timar2017nonbacktracking,allard2019accuracy}. This infinite network consists of an infinite number of copies of the original network, interconnected by edges that replace edges in each copy. It is locally indistinguishable from the original finite network but devoid of cycles. 
Based on the observation that iterating message passing equations on the original finite network and on this infinite network is equivalent, Allard and Hébert-Dufresne~\cite{allard2019accuracy} have argued that message passing can ``see'' the infiniteness of the components with cycles and distinguish them from trees, which remain finite-sized upon translation. 
While this provides a compelling explanation for why message passing solutions often exhibit sharper transitions than simulation results~\cite{timar2017nonbacktracking}, it offers no straightforward answer as to what the upstream object is.

In this paper, we propose a much more intuitive interpretation of the message in the equations above: that it represents the probability that each node is not reachable from any cycle of length greater than two, and the complement represents the probability that it is reachable from multiple cycles. This interpretation is compatible with previous findings, but we believe it provides clearer insights.

Let us begin by solving Eq.~\eqref{eq:mp_simple} on a weakly connected directed graph $G = (V, E)$. If the original network is undirected, we convert it to a directed graph $G$ where each undirected edge is replaced by two reciprocal directed edges. In order to interpret the messages as reachability, neighbors are defined as nodes with outgoing edges to the focal node, i.e., $\partial_i = \{j \mid (j, i) \in E\}$. It is useful to introduce a dual graph of $G$, denoted by $M_G$, that describes the dependencies between messages. The nodes in $M_G$ are the messages, and a directed edge from $x_{k \to j}$ to $x_{j \to i}$ exists if $k \in \partial_{j \to i}$. 

If $M_G$ is a directed acyclic graph, the only solution to Eq.~\eqref{eq:mp_simple} is a trivial one: $x_{j \to i} = 1$ for all messages. This is easy to see. An acyclic graph has at least one node with in-degree zero. When the in-degree of node $x_{j \to i}$ in $M_G$ is zero, it implies $x_{j \to i} = 1$ because there are no messages from $\partial_{j \to i}$ to $j$ and the rhs of Eq.~\eqref{eq:mp_simple} is the product of an empty set. This value-one message, when received by $i$, does not contribute to outgoing messages from $i$; hence, removing such a message does not alter the outcome. By iteratively removing value-one messages (i.e., nodes with in-degree zero in $M_G$), we are eventually left with no messages, meaning all messages have a value of one. We refer to this iterative removal of nodes with in-degree zero as \emph{root removal}.

\begin{figure}
\centering
\includegraphics[width=\linewidth]{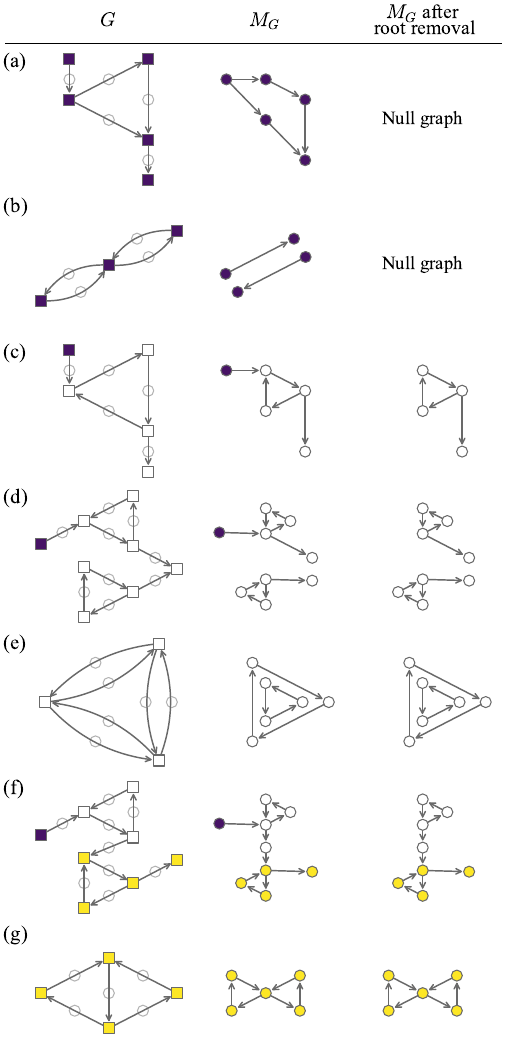}
\caption{Directed graph $G$, corresponding dual graph $M_G$, and $M_G$ after root removal. The nodes in $G$ and $M_G$ are denoted by squares and circles, respectively. The color of each node represents its message or node marginal value: one (dark purple), zero (yellow), and non-convergence (white).}
\label{fig:simple}
\end{figure}

If $G$ is acyclic, $M_G$ is also acyclic [Fig.~\ref{fig:simple}(a)]. In addition, $M_G$ is acyclic if all cycles in $G$ are of length two, that is, consisting of reciprocal edges between two nodes, because the message that is sent from the receiver is discounted [Fig.~\ref{fig:simple}(b)]. This includes the case where the original network is an undirected tree. For Eq.~\eqref{eq:mp_simple} to have a non-trivial solution, $G$ must contain at least one cycle of length greater than two, which we call a \emph{non-reciprocal} cycle.

A non-reciprocal cycle in $G$ translates to a cycle of the same length in $M_G$. Consider $G$ that contains only one such cycle. After root removal to disregard messages that do not contribute to others, each node in $M_G$ that sits on the cycle has exactly one predecessor node. Thus, at each step, each node simply inherits the value of its predecessor from the previous step. The messages therefore circulate in the cycle without modification, causing iteration over Eq.~\eqref{eq:mp_simple} to fail to converge [Fig.~\ref{fig:simple}(c)]. Even when $G$ contains multiple non-reciprocal cycles, non-convergence arises for any cycle in $M_G$ with no incoming edges from outside the cycle after root removal [Fig.~\ref{fig:simple}(d)]. This includes cases where the only cycles in $G$ are two directed cycles that are reversals of each other---for example, when the original network is an undirected unicyclic network [Fig.~\ref{fig:simple}(e)].

A message converges to a non-trivial solution if and only if it is on a cycle in $M_G$ that has at least one incoming edge after root removal, or is reachable from such a cycle. In this case, the message in the cycle at the entry point of an incoming edge is updated at each step by a product of multiple messages whose values are less than one, until it eventually converges to zero. This converged value then propagates to all messages in the cycle and downstream. Note that such an incoming edge after root removal implies the presence of another cycle upstream [Fig.~\ref{fig:simple}(f, g)]. 

To summarize, each message as the solution of Eq.~\eqref{eq:mp_simple} behaves as follows: (i) it converges to one if it is removed by root removal of $M_G$, (ii) it converges to zero if it is reachable from multiple cycles, and (iii) it does not converge otherwise.

The message passing algorithm is often used to compute the marginal value $y_i$ associated with each node $i$. After solving Eq.~\eqref{eq:mp_simple}, node marginals are calculated as
\begin{equation}
y_i = \prod_{j \in \partial_i} x_{j \to i}.
\label{eq:marginal_simple}
\end{equation}
Based on the relationship between $G$ and $M_G$, we can draw a parallel between the behaviors of messages and node marginals. Namely, each node marginal $y_i$ behaves as follows: (i) it converges to one if $i$ is not reachable from any non-reciprocal cycle in $G$, (ii) it converges to zero if $i$ is reachable from multiple non-reciprocal cycles, and (iii) it does not converge if $i$ is reachable from exactly one non-reciprocal cycle. In particular, for an undirected network, the value of $y_i$ signifies whether the component to which $i$ belongs is acyclic, multicyclic, or unicyclic. 

A natural deduction from the above observation is that the messages and node marginals represent the probability that they are not reachable from any cycle in $M_G$ and $G$, respectively, and the complement equals the probability that they are reachable from multiple cycles, as long as the probability of being reachable from exactly one cycle is negligible. We hypothesize that this interpretation extends to percolation models, where messages satisfy Eq.~\eqref{eq:mp_bond} for bond percolation and Eq.~\eqref{eq:mp_site} for site percolation, and node marginals are computed as
\begin{equation}
y_i = \prod_{j \in \partial_i} [1 - q + q x_{j \to i}]
\label{eq:marginal_bond}
\end{equation}
for bond percolation and 
\begin{equation}
y_i = 1 - q + q \prod_{j \in \partial_i} x_{j \to i}
\label{eq:marginal_site}
\end{equation}
for site percolation. 

To test our interpretation, we compare the message passing solutions with empirical probabilities computed from Monte Carlo simulations of percolation with nodes/edges retained with probability $q$.
We use the notation $\hat{p}^\mathrm{A}_i$, $\hat{p}^\mathrm{U}_i$, $\hat{p}^\mathrm{M}_i$, and $\hat{p}^\mathrm{L}_i$ to represent the empirical probabilities that node $i$ is reachable from zero, one, or multiple non-reciprocal cycles, and from the largest (strongly connected) component, respectively, with $\hat{p}^\mathrm{A}_i + \hat{p}^\mathrm{U}_i + \hat{p}^\mathrm{M}_i = 1$. 

\begin{figure}
\centering
\includegraphics[width=\linewidth]{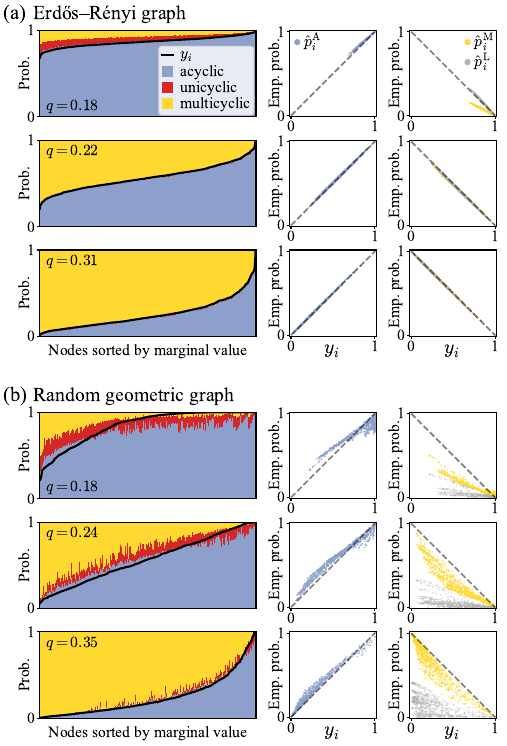}
\caption{Node-wise comparison between node marginal $y_i$, obtained by solving Eqs.~\eqref{eq:mp_bond}, and empirical probabilities $\hat{p}^\mathrm{A}_i$, $\hat{p}^\mathrm{U}_i$, $\hat{p}^\mathrm{M}_i$, and $\hat{p}^\mathrm{L}_i$ of belonging to an acyclic, unicyclic, multicyclic, and the largest component in bond percolation. Empirical probabilities are computed from 5000 independent realizations. The networks are (a) an Erdős--Rényi graph with $n = 1000$ nodes and edge probability $p = 0.006$, and (b) a random geometric graph with $n = 1000$ nodes and connection radius $r = \sqrt{0.006 / \pi}$.}
\label{fig:nodewise}
\end{figure}

\begin{figure}
\centering
\includegraphics[width=\linewidth]{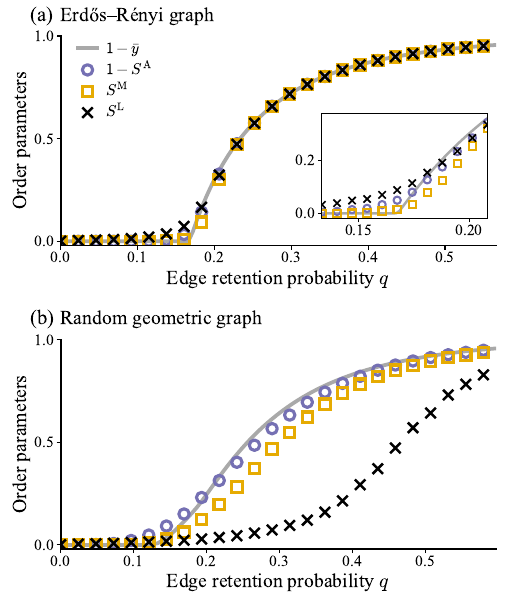}
\caption{Mean node marginal ($\bar{y}$) and the mean fractions of nodes in acyclic components ($S^\mathrm{A}$), in multicyclic components ($S^\mathrm{M}$), and in the largest component ($S^\mathrm{L}$), as a function of edge retention probability $q$ in bond percolation. The fractions are averaged over 500 realizations. The networks are the same as in Fig.~\ref{fig:nodewise}.}
\label{fig:size}
\end{figure}

Figure~\ref{fig:nodewise}(a) shows the comparison of these probabilities for the bond percolation model on Erdős--Rényi graphs with $n$ nodes and edge probability $p$. Deep into the supercritical regime (i.e., $q \gg 1 / (n p)$), $y_i$ and $1 - y_i$ are in excellent agreement with $\hat{p}^\mathrm{A}_i$ and $\hat{p}^\mathrm{M}_i$. In this regime, all components are either acyclic or multicyclic while unicyclic components are extremely rare. However, near the critical point (i.e., $q \approx 1 / (n p)$), unicyclic components become significant, and the marginal values deviate slightly from the empirical probabilities, though the linear correlation remains strong. The agreement between the message passing solutions and empirics is evident at the aggregated level as well. As shown in Fig.~\ref{fig:size}(a), the mean marginal value, $\bar{y} = \sum_i y_i / n$, is consistent with the mean fraction of nodes in acyclic components, $S^\mathrm{A} = \sum_i \hat{p}^\mathrm{A}_i / n$, and multicyclic components, $S^\mathrm{M} = \sum_i \hat{p}^\mathrm{M}_i / n$, such that $1 - \bar{y} \simeq 1 - S^\mathrm{A} \simeq S^\mathrm{M}$ across different edge retention probabilities $q$.

This is anticipated by random graph theory. It has been shown for Erdős--Rényi graphs~\cite{bollobas2001random} and for configuration model graphs~\cite{molloy1995critical} that every component is almost surely a tree or unicyclic in the subcritical regime, and only one component---the giant component---contains multiple cycles in the supercritical regime. Therefore, counting the number of cycles in each component and identifying the giant component are equivalent in these network models. As seen in Fig.~\ref{fig:nodewise}(a), the message passing solutions accurately estimate the probability $\hat{p}^\mathrm{L}_i$ that each node $i$ belongs to the largest component, and consequently, the mean relative size of the largest component, $S^\mathrm{L} = \sum_i \hat{p}^\mathrm{L}_i / n$ [Fig.~\ref{fig:size}(a)].

However, this equivalence cannot be assumed for arbitrary networks. As an illustrative example, let us consider bond percolation on a random geometric graph where $n$ nodes are uniformly distributed on a unit torus, and two nodes are linked if the Euclidean distance between them is smaller than or equal to the connection radius $r$. Figure~\ref{fig:nodewise}(b) shows that the message passing solutions yield reasonable agreement with the empirical probabilities for cyclicity, $\hat{p}^\mathrm{A}_i$ and $\hat{p}^\mathrm{M}_i$, albeit with some discrepancies likely due to the prevalence of unicyclic components. In comparison, it no longer provides a good estimate for the empirical probability $\hat{p}^\mathrm{L}_i$ of being in the largest component. This is because many components can still contain multiple cycles despite not being the largest due to the abundance of short cycles in the original network. As a consequence, the message passing solution aligns much more closely with the mean number of nodes in acyclic and multicyclic components than with the mean size of the largest component [Fig.~\ref{fig:size}(b)].

\begin{figure}
\centering
\includegraphics[width=\linewidth]{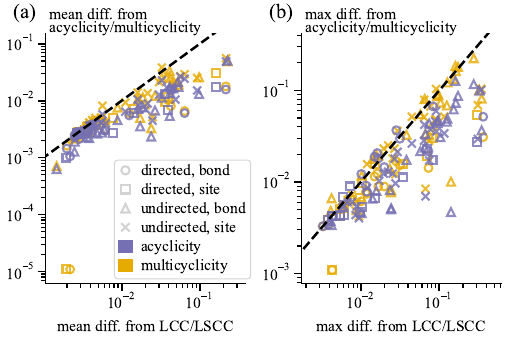}
\caption{Deviation of the message-passing solutions (node marginals) from the empirical probabilities for bond or site percolation on real-world networks. For each network and each value of $q$ between 0 and 1 in increments of 0.01, the empirical probabilities for each node are estimated from $10^4$ realizations. The differences in probability are averaged over all nodes for each $q$, and aggregated across $q$ by taking the mean (a) or the maximum (b). The dashed line denotes identity.}
\label{fig:diff}
\end{figure}

We further examine the agreement between the message passing solution and the empirical probabilities for bond and site percolation on 43 real-world networks (27 undirected and 16 directed; see Supplemental Material for details). 
For each value of node/edge retention probability $q$, we calculate the mean absolute differences over all nodes as $\sum_i \left|y_i - \hat{p}^\mathrm{A}_i\right| / n$, $\sum_i \left|1 - y_i - \hat{p}^\mathrm{M}_i\right| / n$, and $\sum_i \left|1 - y_i - \hat{p}^\mathrm{L}_i\right| / n$. We then aggregate them across $q$ by computing their mean or maximum. Figure~\ref{fig:diff} shows that the difference between the message passing solution and the empirical probabilities for cyclicity is on par with or smaller, in some cases by more than an order of magnitude, than the difference between the message passing solution and the empirical probability of being reachable from the largest component. This corroborates our claim that the message passing algorithm identifies the cyclicity, rather than the extensivity, of the components that can reach each node. 

In conclusion, we have shown that message passing for percolation essentially captures cyclicity---more precisely, it responds to the number of cycles that reach each node. This interpretation is generally valid for bond and site percolation on directed and undirected networks alike. The ability to identify the largest component is not an intrinsic feature of the algorithm but rather a byproduct. The prediction is accurate when the largest component is the only one with multiple cycles, but performance deteriorates when cyclicity is decoupled from component size. 
We note that neither sparsity nor locally treelikeness guarantees accuracy: sparse networks can be rich in short cycles (e.g., random geometric graphs), and locally treelike networks can have more than one component with multiple cycles (e.g., stochastic block model graphs).

Our findings imply that percolation processes undergo two distinct structural transitions: the emergence of the giant component and the transition in cyclicity. These two transitions have long been conflated because they happen to coincide asymptotically for Erdős--Rényi graphs and configuration model networks, and approximately for a broad class of networks. In fact, this coincidence has been explicitly exploited to develop an efficient heuristic for network dismantling~\cite{braunstein2016network,zdeborova2016fast}, and, as our results suggest, serves as an implicit prerequisite for message passing algorithms to identify the giant component. Yet, these two transitions are separable in general, which underscores the need to recognize the transition in cyclicity as a distinct phenomenon that merits independent investigation.

\begin{acknowledgments}
The author thanks Lasse Leskelä for helpful discussions. This study was supported by the Strategic Research Council established within the Research Council of Finland (decision numbers 364386 and 364371). Some of the calculations in this study were performed using computational resources provided by the Aalto Science-IT project.

All real-world network datasets used in this study are publicly available from the Netzschleuder repository~\cite{peixoto2020netzschleuder}. Details about the datasets are given in the Supplemental Material. 
\end{acknowledgments}

%

\end{document}


\title{Supplemental Material for \\``Message Passing and Cyclicity Transition''}
\author{Takayuki Hiraoka}
\email{takayuki.hiraoka@aalto.fi}
\affiliation{Department of Computer Science, Aalto University, 00076 Espoo, Finland} 

\maketitle

\section{Real-world network datasets}
Table~\ref{tab:networks} summarizes the real-world network datasets used in this study. The datasets were obtained from the Netzschleuder repository \cite{peixoto2020netzschleuder} (accessed on 2026-03-18), which provides a large collection of network datasets from various domains. Each dataset is identified by a unique Netzschleuder identifier, and the table includes a brief description, reference to the primary source, number of nodes ($N$), number of edges ($E$), and whether the network is directed or undirected. When a dataset contains multiple networks, the dataset identifier and the name of the network is separated by a slash, e.g., \texttt{arxiv\_collab/cond-mat-1999}.

\begin{longtable}{lllrrl}
\caption{Real-world network datasets used in this study.}
\label{tab:networks} \\
\toprule
\textbf{Netzschleuder identifier} & \textbf{Description} & \textbf{Ref.}
  & \textbf{$N$} & \textbf{$E$} & \textbf{Type} \\
\midrule
\endfirsthead
\multicolumn{6}{c}{\tablename\ \thetable{} (continued)} \\
\toprule
\textbf{Netzschleuder identifier} & \textbf{Description} & \textbf{Ref.}
  & \textbf{$N$} & \textbf{$E$} & \textbf{Type} \\
\midrule
\endhead
\midrule
\multicolumn{6}{r}{\textit{Continued on next page}} \\
\endfoot
\bottomrule
\endlastfoot
\texttt{adjnoun} & Word adjacencies of David Copperfield & \cite{newman_finding_2006} & 112 & 425 & Undirected \\
\texttt{arxiv\_collab/cond-mat-1999} & Scientific collaborations in physics (1995-2005) & \cite{newman_structure_2001} & 16726 & 47594 & Undirected \\
\texttt{arxiv\_collab/hep-th-1999} & Scientific collaborations in physics (1995-2005) & \cite{newman_structure_2001} & 8361 & 15751 & Undirected \\
\texttt{board\_directors/net1m\_2011-08-01} & Norwegian Boards of Directors (2002-2011) & \cite{seierstad_few_2011} & 1421 & 3855 & Undirected \\
\texttt{celegans\_metabolic} & Metabolic network (C. elegans) & \cite{duch_community_2005} & 453 & 4596 & Undirected \\
\texttt{cora} & CORA citations (1998) & \cite{mccallum_automating_2000} & 23166 & 91500 & Directed \\
\texttt{dolphins} & Dolphin social network (1994-2001) & \cite{lusseau_bottlenose_2003} & 62 & 159 & Undirected \\
\texttt{drosophila\_flybi} & Fruit fly protein interactions (Drosophila melanogaster) & \cite{tang_nextgeneration_2023} & 2939 & 8723 & Undirected \\
\texttt{ecoli\_transcription/v1.1} & E. coli transcription network (2002) & \cite{shen-orr_network_2002} & 423 & 578 & Directed \\
\texttt{euroroad} & Euroroad network (2011) & \cite{subelj_robust_2011} & 1174 & 1417 & Undirected \\
\texttt{faa\_routes} & FAA Preferred Routes (2010) & \cite{kunegis_konect_2013} & 1226 & 2615 & Directed \\
\texttt{facebook\_friends} & Maier Facebook friends (2014) & \cite{maier_cover_2017} & 362 & 1988 & Undirected \\
\texttt{facebook\_organizations/L1} & Within-organization Facebook friendships (2013) & \cite{fire_organization_2016} & 5793 & 45266 & Undirected \\
\texttt{facebook\_organizations/L2} & Within-organization Facebook friendships (2013) & \cite{fire_organization_2016} & 5524 & 94219 & Undirected \\
\texttt{foodweb\_little\_rock} & Little Rock Lake food web (1991) & \cite{martinez_artifacts_1991} & 183 & 2494 & Directed \\
\texttt{football\_tsevans} & NCAA college football 2000 (corrected metadata) & \cite{girvan_community_2002,evans_clique_2010} & 115 & 613 & Undirected \\
\texttt{gnutella/04} & Gnutella p2p networks (2002) & \cite{ripeanu_mapping_2002} & 10879 & 39994 & Directed \\
\texttt{gnutella/06} & Gnutella p2p networks (2002) & \cite{ripeanu_mapping_2002} & 8717 & 31525 & Directed \\
\texttt{gnutella/09} & Gnutella p2p networks (2002) & \cite{ripeanu_mapping_2002} & 8114 & 26013 & Directed \\
\texttt{gnutella/25} & Gnutella p2p networks (2002) & \cite{ripeanu_mapping_2002} & 22687 & 54705 & Directed \\
\texttt{google} & Google internal webpages (2007) & \cite{palla_directed_2007} & 15763 & 171206 & Directed \\
\texttt{interactome\_figeys} & Figeys human interactome (2007) & \cite{ewing_largescale_2007} & 2239 & 6452 & Directed \\
\texttt{interactome\_stelzl} & Stelzl human interactome (2005) & \cite{stelzl_human_2005} & 1706 & 6207 & Directed \\
\texttt{interactome\_vidal} & Vidal human interactome (2005) & \cite{rual_proteomescale_2005} & 3133 & 6726 & Undirected \\
\texttt{interactome\_yeast} & Coulomb yeast interactome (2005) & \cite{coulomb_gene_2005} & 1870 & 2277 & Undirected \\
\texttt{internet\_as} & Internet AS graph (2006) & \cite{karrer_percolation_2014} & 22963 & 48436 & Undirected \\
\texttt{jazz\_collab} & Jazz collaboration network & \cite{gleiser_community_2003} & 198 & 2742 & Undirected \\
\texttt{karate/78} & Zachary Karate Club & \cite{zachary_information_1977} & 34 & 78 & Undirected \\
\texttt{marvel\_universe} & Marvel Universe social network & \cite{alberich_marvel_2002} & 19428 & 95497 & Undirected \\
\texttt{polblogs} & Political blogs network (2004) & \cite{adamic_political_2005} & 1490 & 19090 & Directed \\
\texttt{polbooks} & Political books network (2004) & \cite{krebs_polbooks} & 105 & 441 & Undirected \\
\texttt{power} & Western US Power Grid & \cite{watts_collective_1998} & 4941 & 6594 & Undirected \\
\texttt{reactome} & Joshi-Tope human protein interactome (2005) & \cite{joshi-tope_reactome_2005} & 6327 & 147547 & Undirected \\
\texttt{route\_views/19991231} & Route Views AS graphs (1997-1998) & \cite{leskovec_graphs_2005} & 2107 & 4676 & Undirected \\
\texttt{route\_views/20000102} & Route Views AS graphs (1997-1998) & \cite{leskovec_graphs_2005} & 6474 & 13895 & Undirected \\
\texttt{terrorists\_911} & 9-11 terrorist network & \cite{krebs_mapping_2002,krebs_uncloaking_2002} & 62 & 152 & Undirected \\
\texttt{ugandan\_village/friendship-16} & Ugandan village networks (2013) & \cite{chami_social_2017} & 372 & 1475 & Undirected \\
\texttt{ugandan\_village/friendship-4} & Ugandan village networks (2013) & \cite{chami_social_2017} & 320 & 2302 & Undirected \\
\texttt{ugandan\_village/friendship-8} & Ugandan village networks (2013) & \cite{chami_social_2017} & 369 & 1902 & Undirected \\
\texttt{uni\_email} & Email network (Uni. R-V, Spain, 2003) & \cite{guimera_selfsimilar_2003} & 1133 & 10903 & Directed \\
\texttt{word\_adjacency/french} & Word Adjacency Networks & \cite{milo_superfamilies_2004} & 8325 & 24295 & Directed \\
\texttt{word\_adjacency/spanish} & Word Adjacency Networks & \cite{milo_superfamilies_2004} & 11586 & 45129 & Directed \\
\texttt{yeast\_transcription} & Yeast transcription network (2002) & \cite{milo_network_2002} & 916 & 1094 & Directed \\
\end{longtable}